\documentclass[onecolumn,superscriptaddress,notitlepage,nofootinbib]{revtex4-1}
\usepackage{epsfig}
\usepackage{amsfonts}
\usepackage{amsmath}
\usepackage{amssymb}
\usepackage{soul}
\graphicspath{{.}{./Figures/}}

\usepackage{color} 
  \newcommand{\beq}{\begin{equation}}
  \newcommand{\eeq}{\end{equation}} 
  \def\nuc#1#2{\relax\ifmmode{}^{#1}{\protect\text{#2}}\else${}^{#1}$#2\fi}
  \def\itnuc#1#2{\setbox\@tempboxa=\hbox{\scriptsize\it #1}
    \def\@tempa{{}^{\box\@tempboxa}\!\protect\text{\it #2}}\relax
    \ifmmode \@tempa \else $\@tempa$\fi}
  
\usepackage{hyperref}

\newcommand{\simle}{\hspace*{0.2em}\raisebox{0.5ex}{$<$}
     \hspace{-0.8em}\raisebox{-0.3em}{$\sim$}\hspace*{0.2em}}

\newcommand{\mrm}[1]{\mathrm{#1}}

\newcommand{\klo}{\ensuremath{k_\mathrm{lo}}}
\newcommand{\khi}{\ensuremath{k_\mathrm{hi}}}
\newcommand{\kC}{\ensuremath{k_\mathrm{C}}}
\newcommand{\Ac}{\ensuremath{A_\mathrm{c}}}
\newcommand{\mR}{\ensuremath{m_\mathrm{R}}}
\newcommand{\pp}{\ensuremath{\boldsymbol{p}}}
\newcommand{\QQ}{\ensuremath{\boldsymbol{Q}}}
\newcommand{\rr}{\ensuremath{\boldsymbol{r}}}
\renewcommand{\d}{\ensuremath{\mathrm{d}}}

\begin{document}

\title{Finite-Size Effects in Heavy Halo Nuclei from Effective Field Theory}
\author{E. Ryberg}
\author{C. Forss\'en}
\affiliation{Department of Physics, Chalmers University of Technology,
  412 96 G\"oteborg, Sweden}
\author{D.~R. Phillips}
\affiliation{Institute of Nuclear and Particle Physics and  Department of Physics \& Astronomy. Ohio University,  Athens, OH 45701, USA}
\affiliation{Institut f\"ur Kernphysik,  Technische Universit\"at
  Darmstadt, 64289 Darmstadt, Germany}
\affiliation{ExtreMe Matter Institute EMMI, GSI Helmholtzzentrum
  f{\"u}r Schwerionenforschung GmbH, 64291 Darmstadt, Germany}
\author{U. van Kolck}
\affiliation{Institut de Physique Nucl\'eaire, CNRS/IN2P3,
  Universit\'e Paris-Sud, Universit\'e Paris-Saclay, 91406 Orsay Cedex,
  France} 
\affiliation{Department of Physics, University of Arizona, Tucson, AZ
  85721, USA} 
\begin{abstract}
Halo/Cluster Effective Field Theory describes halo/cluster nuclei in
an expansion in the small ratio of the size of the core(s) to the size
of the system.  Even in the point-particle limit, neutron halo nuclei
have a finite charge radius, because their center of mass does not
coincide with their center of charge.  This point-particle
contribution decreases as $1/\Ac$, where $\Ac$ is the mass number of
the core, and diminishes in importance compared to other effects,
e.g., the size of the core to which the neutrons are bound. Here we
propose that for heavy cores the EFT expansion should account for the
small factors of $1/\Ac$.  As a specific example, we discuss the
implications of this organizational scheme for the inclusion of
finite-size effects in expressions for the charge radii of halo
nuclei. We show in particular that a short-range operator could be the dominant
effect in the charge radius of one-neutron halos bound by a P-wave
interaction.  The point-particle contribution remains the leading
piece of the charge radius for one-proton halos, and so Halo EFT has
more predictive power in that case.
\end{abstract}

\maketitle

\section{Introduction
\label{intro}}
%
Theoretical models of many-body systems usually treat the constituent
particles as having no internal structure. This point-particle approximation is
also used in cluster models, e.g., for the description of halo
systems~\cite{Zhukov:1993aw,Hansen:1995pu,Jensen:2004zz,Frederico:2012xh}, 
even though
one might encounter situations for which the core is rather large.
Finite-size effects are included \emph{a posteriori}, but can become
significant for certain observables.  As an example, the total charge
radius is usually calculated by simply adding in quadrature~\cite{Friar:1975pp}
the charge radii of the constituents and the calculated point-particle
radius, see e.g.\ the calculation of charge radii for neutron-rich
helium isotopes in the Gamow Shell
Model~\cite{Papadimitriou:2011jx}. Instead, it would be useful to construct a
framework in which finite-size effects can be included
systematically. 

Constituent-size effects can be accounted for in effective field theories
(EFTs), where they appear through derivative interactions.
For example, the nucleon charge radius
(and, more generally, nucleon form factors) can be calculated 
\cite{Bernard:1995dp}
in Chiral Perturbation Theory (ChPT) in an expansion
in powers of $k_{\pi}/M_{\rm QCD}$, where $k_{\pi}\sim 150~\mrm{MeV}$ 
is a momentum scale associated with the lightest carrier
of the nuclear force, the pion, and
$M_{\rm QCD}\sim 1$ GeV is the characteristic mass scale of QCD.
The relevant pion parameters 
are its mass and decay constant.
Chiral EFT~\cite{Bedaque:2002mn} 
is a generalization of ChPT to a typical nucleus,
for which the binding energy per nucleon is $B/A\sim k_\pi^2/M_{\rm QCD}$
and the radius $R\sim A^{1/3}/k_\pi$.
The nuclear charge radius includes the sum of the
nucleons' radii plus many-body effects generated by 
internucleon interactions and currents \cite{Phillips:2016mov}.
We would like to have a similar framework for clusterized
systems.

Clusterized systems, with much smaller energies and larger radii, are
additionally characterized by scales beyond the pion scales. These
nuclei can be viewed as a collection of valence nucleons orbiting
around either no core (few-nucleon systems), one core (halo nuclei) or many
cores (cluster nuclei). The cores themselves frequently---but not
always---have properties of typical nuclei.  The generic existence of
such systems can be understood as a consequence of a fine-tuning in
QCD, which introduces a lighter momentum scale $\aleph \sim 30$ MeV
\cite{Beane:2001bc,vanKolck:2008bm}.  For such loosely bound systems
we can devise EFTs that exploit the separation of scales without
involving pions explicitly.  In these EFTs one considers processes
with typical momenta $k_\mrm{lo}$, such that
$k_\mrm{lo}\ll k_\mrm{hi}$, where $k_\mrm{hi}\simle k_\pi$ is a
high-momentum scale. One then develops an expansion for observables in
powers of $k_\mrm{lo}/k_\mrm{hi}$.  The very lightest nuclei are
dilute systems with no core, where the dominant (two- and
three-nucleon) interactions are S-wave.  The corresponding EFT is
Pionless EFT, for which power counting is relatively well understood
\cite{Bedaque:2002mn}.

Halo/Cluster EFT, here labeled HEFT, was proposed 
as an EFT for systems with one \cite{Bertulani:2002sz,Bedaque:2003wa} 
or more \cite{Higa:2008dn} cores and valence 
nucleons~\cite{Canham:2008jd,Canham:2009xg}. 
(See {R}ef.~\cite{Hammer:2017tjm} for a recent review.)
HEFT power counting is a generalization of the power counting
for Pionless EFT allowing for
dominant interactions in waves with non-vanishing angular momentum
and for a breakdown scale $k_\mrm{hi}$ estimated from the first excitation
of the core and/or its size.
In the first cases considered, 
$^{5,6}$He \cite{Bertulani:2002sz,Bedaque:2003wa,Rotureau:2012yu,Ji:2014wta,Ryberg:2017tpv}, 
there is an alpha-particle core, and the neutron-alpha ($n$-$\alpha$) 
interaction is mostly of 
P-wave nature, generating a near-threshold $^{5}$He resonance.
The $\alpha$-$\alpha$ interaction, in turn, is obtained \cite{Higa:2008dn} from 
$\alpha$-$\alpha$ scattering and the lowest $^8$Be state.
HEFT has since been extended to heavier cores
and to 
proton halo 
systems~\cite{Hammer:2017tjm}.
In most of these cases the high scale $k_\mrm{hi}$ in HEFT is associated with 
the size of the core, i.e.
$k_\mrm{hi} \sim 1/R_\mrm{c}$. This means HEFT is an expansion in powers of
$R_\mrm{c}/R_\mrm{halo}$, where $R_\mrm{halo}$ is the unnaturally large size
 of the halo system. 
 
The different contributions to the charge radius of a halo nucleus can then
be organized in the HEFT expansion and the size of the effect due to the 
finite size of the core (and of the nucleon) estimated. We do that and thereby 
derive---for both S- 
and P-wave one-neutron halos---the charge radius formula that is frequently 
used in nuclear theory. However, we also point out that there is, in principle,
another expansion parameter 
present when HEFT is applied to systems with a 
relatively large number, $\Ac$, of core nucleons. 
To leading order in $1/\Ac$ 
the core is static, its recoil being
small compared to nucleon recoil.
Consequently the center of mass of a neutron halo coincides with its center 
of charge.
Thus, whereas for light halos (e.g. ${}^6$He) the difference between these two 
generates
an important contribution to the charge radius~\cite{Lu:2013ena}, 
for heavier systems
the corresponding effect goes to zero.  Correctly assessing the impact of the
finite size of the constituents on the charge radius requires keeping track of 
factors of 
$1/\Ac$. 

Moreover, the charge radii of halo nuclei are affected by a short-range 
operator, which is subleading in $R_{\mrm c}/R_\mrm{halo}$ but leading in $1/\Ac$. 
We show that for one-neutron halos bound by a P-wave interaction (e.g. 
the excited state of \nuc{11}{Be}) this
effect may be as important as the long-distance contributions to the halo's 
charge radius that have been previously computed in HEFT~\cite{Hammer:2011ye}.
Similar considerations also apply to the form factors of
two-neutron halos such as those discussed in {R}ef. \cite{Vanasse:2016hgn}.
They are not, however, as pressing for proton halos, where a finite charge 
radius will be generated by the photon coupling 
to the valence proton(s) even if the core is infinitely heavy. 

This exemplifies the importance of keeping track---to the extent possible---of 
factors of $1/A_\mrm{c}$ in observables, rather than just counting powers of 
$k_\mrm{lo}$. 
This is quite similar to the need to distinguish between relativistic 
corrections that are suppressed
by powers of the inverse nucleon mass, $1/m_\mrm{N}$, and other corrections that 
only carry powers of 
$1/k_\mrm{hi}$~\cite{vanKolck:1998bw,Chen:1999tn,Zhang:2017yqc}. 
The significant difference between these two scales
produces a hierarchy between effects that scale with the same power of 
$k_\mrm{lo}$. 
Only once that hierarchy has been identified can the power counting be 
formulated in an efficient manner. 

We isolate the $\Ac$ dependence that enters the charge radius through 
kinematic effects, i.e., because the nucleon-core mass ratio,
$m_\mrm{N}/m_\mrm{c}$, is small. 
In contrast, we assume that all Lagrangian coefficients (LECs) scale with 
a power of $k_\mrm{hi}$ that is solely determined by the naive engineering 
dimension of the operator they multiply, i.e., 
we use naive dimensional analysis with respect to $k_\mrm{hi}$ and do not 
attach any additional $\Ac$ dependence to the LECs. It is true that 
$k_\mrm{hi}$ is also $\Ac$ dependent, because $k_\mrm{hi}$ will generically be 
of order the inverse core size, $1/R_\mrm{c}$, 
and $R_\mrm{c}$ can be taken to be $\propto \Ac^{1/3}$. 
But any additional accounting of the $\Ac$ dependence of short-distance physics
in HEFT beyond this would require a more microscopic
understanding of the $\Ac$ dependence of all the cofficients in the EFT. 
This could be achieved by matching HEFT to 
a microscopic calculation, but it could be argued that such matching 
goes beyond the EFT philosophy of writing down a theory that is independent 
of the short-distance physics. 
In contrast, the kinematic effects we identify here are universal, 
in the sense that they occur irrespective of the nature of the forces between
the halo nucleons and the core(s). 

Here we focus on the 
charge radius of single-neutron halo nuclei, where the point contribution 
is suppressed by $1/\Ac$ for the reason described above. But, the
presence of the heavy core propagator is ubiquitous, so similar effects 
will affect other 
observables as well. For example, one expects the Born-Oppenheimer 
approximation to emerge in 
systems with multiple heavy cores and/or valence nucleons. 

Our paper is structured as follows: 
In Sec.~\ref{power} we discuss the interplay between the various
scales that are involved in an EFT for a halo system with  a heavy
core. We also discuss the low-energy scattering parameters for the nucleon-core
system and introduce the charge radius in terms of the
momentum expansion of the low-energy charge form factor.
In Sec.~\ref{sec:3} we derive the observable charge radius for S-
and P-wave one-neutron halo states. The
power counting is exemplified by considering the charge radius for selected halo
states. We summarize our findings in Sec.~\ref{conclusion}. An appendix
discusses the corresponding results for proton halos, where considering factors
of $1/\Ac$ does
not lead to any change in the  hierarchy of the various physical mechanisms
that contribute to the charge radius.

\section{Power Counting}
\label{power}
%
Once the relevant degrees of freedom are chosen,
a model consists of a specific set of interactions among them.
In contrast, with an EFT one considers the most general dynamics
consistent with the known symmetries. It is crucial to
organize the corresponding infinity of contributions
to any observable according to their size (``power counting'').

We are interested in a clusterized system where the
size $R_\mrm{halo}\sim 1/k_{\rm lo}$ of the system is sufficiently large 
that the constituents can be taken as point-like in a first approximation.
This system might be probed with particles 
(photons, electrons, neutrinos, nucleons) of wavelength $\sim 1/k_{\rm lo}$ 
that cannot resolve the inner structure of the constituents.
For simplicity we consider a few valence nucleons 
orbiting around 
a single core of radius $R_\mrm{c}$ 
consisting of $\Ac\gg 1$ nucleons. 
The arguments below can be generalized straightforwardly to multiple-core 
systems.

The power counting of HEFT is based, like that of other EFTs, on the 
ratio of momentum scales, $k_\mrm{lo}/k_\mrm{hi}\ll 1$.
The high momentum scale $k_\mrm{hi}$ is determined by physics not accounted
for explicitly in HEFT.
Since nuclei are bigger than nucleons we must 
have 
$k_\mrm{hi}\simle 1/R_\mrm{N}\sim k_\pi$, with $R_\mrm{N}$ the size of a nucleon, which
is generically set by pionic dynamics described by ChPT. But
a more restrictive condition on 
$k_\mrm{hi}$ arises from the requirement that 
details of the core are not resolved:
\begin{equation}
k_\mrm{hi}\simle 1/R_\mrm{c}.
\label{khi}
\end{equation}
Adopting the standard rule for the scaling of the nuclear size with
$\Ac$ we then have $k_\mrm{hi} \sim \Ac^{-1/3} k_\mrm{\pi}$, although we
note that several of the cores considered up until now in HEFT are
somewhat larger than this valley-of-stability lore indicates.
If there exist low-lying excited states of the core, corresponding to
a lower momentum scale than the inverse size of the core, the
high-momentum scale needs to be adjusted accordingly, or else explicit
degrees of freedom must be introduced for the low-lying excited
states. One example of this is in the $^7{\rm Be}(p,\gamma){}^8{\rm B}$ 
reaction,
where the low-lying excited state of ${}^7$Be must be included as an explicit
degree of freedom in the HEFT if Eq.~\eqref{khi} is to 
prevail~\cite{Zhang:2013kja,Zhang:2014zsa}.  

$k_\mrm{hi}^{-1}$ can therefore be expected to increase for heavier systems,
rendering finite-size effects more important. Meanwhile,
$1/A_\mrm{c}$ decreases and 
will generically be smaller than $k_\mrm{lo}/k_\mrm{hi}$.
It introduces an additional expansion parameter.
Explicit factors of $1/A_\mrm{c}\ll 1$ enter through the core mass,
\begin{equation}
m_\mrm{c} \approx A_\mrm{c} m_\mrm{N},
\label{eq:coremass}
\end{equation}
where the average nucleon mass $m_\mrm{N}\simeq 940$ MeV and
 we have neglected the effect of nuclear binding for the kinematic
 purposes we have in mind here. 

In our non-relativistic theory approximate Galilean invariance
guarantees that the interactions depend on the mass of the particles
only in trivial ways that can be scaled out of the theory.
Therefore all explicit occurrences of $1/\Ac$ are associated with the 
kinematics of the 
two-particle system, and once again, Galilean invariance means that they must 
be encoded in the halo's reduced mass, 
\begin{equation}
\mR=\frac{m_\mrm{N}m_\mrm{c}}{M_\mrm{cN}}
=m_\mrm{N}\left(1-\frac{1}{A_\mrm{c}}+\ldots \right) ,
\label{cNredmass}
\end{equation}
where
\begin{equation}
M_\mrm{cN}= m_\mrm{c}+m_\mrm{N}
\equiv \frac{m_\mrm{N}}{f},
\label{cNtotmass}
\end{equation}
is the total nucleon-core mass, with $f\simeq1/(\Ac+1)\sim1/\Ac$.
Specifically, $m_\mrm{R}$ has a fractional
difference from $m_\mrm{N}$ of $\approx 1/\Ac$, reflecting the extent to
which the core is still dynamical in the (effective) one-body problem.

The large mismatch in masses evident in Eq. \eqref{eq:coremass} means that
the pertinent energy scale for the valence nucleon is the one-nucleon 
separation energy 
\begin{equation}
B_\mrm{s} \sim \frac{k_\mrm{lo}^2}{2 m_\mrm{R}},
\label{EN}
\end{equation}
which is much smaller than the binding energy of the core but
much larger than the recoil energy of the core
\begin{equation}
E_\mrm{c} \sim B_\mrm{s}/A_\mrm{c}.
\label{eq:Ec}
\end{equation}
For energies of the order of the typical nucleon energy,
$E\sim B_\mrm{s}$, nucleon recoil is a leading-order effect. Beyond leading order
the ratio $k_\mrm{lo}/m_\mrm{N}$ occurs only in (small) relativistic corrections.
In contrast, core recoil is suppressed by a factor of $1/A_\mrm{c}$.
Thus at leading order (LO) the core propagator is static, that is,
\begin{equation}
S_\mrm{c}(E,\pp)=\frac{1}{E-\frac{\pp^2}{2m_\mrm{c}}+i\varepsilon}
\to S_\mrm{c}^\mrm{LO}(E)=\frac{1}{E+i\varepsilon}.
\end{equation}
Thus, for low-order calculations one can simplify the procedure by using 
a static core, and including recoil effects perturbatively
as higher-order corrections. Relativistic corrections that scale like 
$k_\mrm{lo}/m_\mrm{c}$ will be even smaller.

In addition to these kinematic factors of $1/A_\mrm{c}$,
there may be dependence on $A_\mrm{c}$ coming through the interaction
coefficients, or ``low-energy constants'' (LECs).
As a trivial example, electromagnetic interactions add up
constructively for protons and the corresponding LECs
in general depend on the core charge $Z_\mrm{c}=A_\mrm{c} - N_\mrm{c}$.
It is not clear how to deal with this quantity {\it a priori}.
In neutron halos, $Z_\mrm{c}$ can be significantly smaller
than $A_\mrm{c}/2$, but this is not necessarily so for proton halos.
We will keep factors of $Z_\mrm{c}$ explicit and deal with them
on a case-by-case basis.

Likewise, the LECs for strong interactions might in specific 
cases represent some constructive or destructive
interference in the interactions of the valence nucleon with
the core nucleons. 
One way to determine the $A_\mrm{c}$ dependence 
of these LECs
is by matching HEFT to the {\it ab initio} solution of the same
system with a more fundamental 
EFT \cite{Zhang:2013kja,Zhang:2014zsa,Hagen:2013jqa}, in a region where 
both EFTs are valid. Another way is to look at systematic trends in LECs fitted
to data for different cores. 
In either case a manifestation of strong $A_\mrm{c}$ dependence would be
a particularly large or small LEC value with respect to
the expected power of $k_\mrm{hi}$. 
Since there is no clear case at this point, 
below we limit ourselves to the kinematical factors arising from the
core mass, although the counting of factors of $1/A_{\rm c}$ could be improved 
later if needed.

\subsection{Nucleon-core scattering}
\label{Nucleoncorescattering}
%
EFTs incorporate from the start the coupling to the continuum,
so that most calculations of halo structure, including
form factors, are intimately connected with nucleon-core scattering.
A discussion of nucleon-core interactions is therefore necessary for the 
calculation of
form factors, and we briefly review previous work on the subject here.

First we consider a halo system where the dominant  
core-nucleon interaction is S-wave.
In this case, an EFT where all forces are of contact type reduces to 
the effective range expansion (ERE) \cite{Bedaque:2002mn,Beane:2000fx}.
One can think of the scattering length $a_0$
as what characterizes the size of the halo system,
and 
the effective range $r_0$ (and higher ERE parameters)
as reflecting the breakdown scale, namely the
size of the core.
The large size of the halo system is manifest in
a large scattering length, while higher effective-range parameters 
are assumed to have sizes set by $1/k_\mrm{hi}$:
\begin{equation}
\begin{split}
a_0\sim1/k_\mrm{lo} \sim R_\mrm{halo} \, , \\
r_0\sim1/k_\mrm{hi}\sim R_\mrm{c} .
\label{a0r0magnitude}
\end{split}
\end{equation}
For an S-wave nucleon bound to the core with separation energy $B_{\mrm{s}0}>0$,
the nucleon-core $T$ matrix has a pole at $k=i\gamma_0$ with
\begin{equation}
\gamma_0\equiv \sqrt{2\mR B_{\mrm{s}0}} \sim k_\mrm{lo}.
\label{gamma0}
\end{equation}

The power counting for this system is almost 
identical to that of Pionless EFT for 
an S-wave bound state \cite{Beane:2000fx,Bedaque:2002mn}, which was adopted,
for example, 
in the form-factor calculation of {R}ef.~\cite{Hammer:2011ye}.
Note, however, that the suppression of the core recoil by 
Eq. \eqref{eq:Ec} 
means that in LO the nucleon-core reduced mass 
that enters scattering is $m_\mrm{N}$, see Eq. \eqref{cNredmass}.

Similar considerations apply to higher partial waves,
but differences arise from the different renormalization:
the more singular character of the interactions requires
more LECs at any given order.
For P-wave nucleon-core scattering, for example,
both the scattering ``length'' $a_1$ and the effective ``range'' $r_1$
appear at LO \cite{Bertulani:2002sz,Bedaque:2003wa}
\footnote{Note that the P-wave scattering
length and effective range have dimensions of volume and
momentum, respectively.}.
The mildest assumption \cite{Bedaque:2003wa} 
is that the effective range, just
as for S waves, is not fine tuned and directly
reflects the breakdown scale,
\begin{equation}
r_1\sim k_\mrm{hi}\sim 1/R_\mrm{c}.
\label{r1magnitude}
\end{equation}
In this case $r_1k^2$ is larger than the unitarity term $ik^3$,
and $S$-matrix poles of non-zero energy require a single fine tuning,
\begin{equation}
a_1\sim 1/(k_\mrm{lo}^2k_\mrm{hi}).
\label{a1magnitude}
\end{equation}
Assuming the higher ERE parameters still scale with $k_\mrm{hi}$,
they are all subleading, and at LO there are two poles: depending
on the sign of $a_1r_1$,
a resonance on the real axis or a real/virtual bound-state pair with 
binding momentum
\begin{equation}
\gamma_1\equiv \sqrt{2\mR B_{\mrm{s}1}} \sim k_\mrm{lo},
\label{gamma1}
\end{equation}
in terms of the separation energy $B_{\mrm{s}1}>0$.
Thus, again, $1/k_\mrm{lo}$ 
is related to the size of the halo system.
At NLO the unitarity term needs to be included.
If treated exactly, a third pole appears
with momentum $\sim k_\mrm{hi}$, that is, outside the EFT.
In the unlikely case where there are three low-energy poles
in the low-energy region, the $k_\mrm{hi}$ in 
Eqs. \eqref{r1magnitude} and \eqref{a1magnitude} should
be replaced by $k_\mrm{lo}$ \cite{Bertulani:2002sz}.

Just as for S waves, the assumption that no further powers of
$1/A_\mrm{c}$ appear in the ERE parameters implies that the
only change in power counting when
treating $1/A_\mrm{c}$ as small is the extra expansion \eqref{cNredmass}.

\subsection{Charge form factor}
\begin{figure}
  \begin{center}
    \includegraphics[scale=0.5]{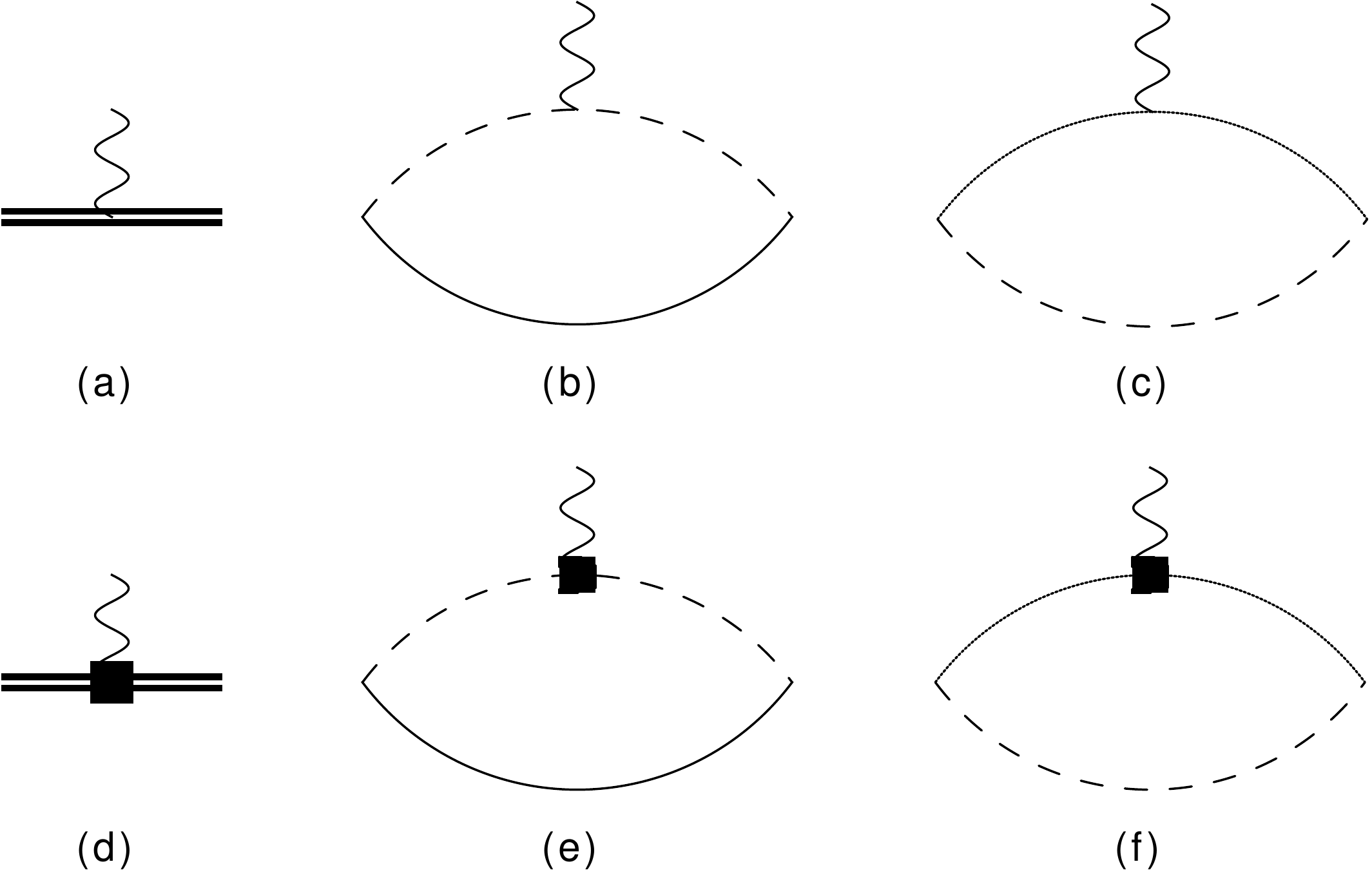}
    \caption{Charge form-factor diagrams for a nucleon-system
      core. A solid/dashed/double/wavy 
    line denotes a nucleon/core/dicluster/photon field.  
    An unmarked (solid square) photon vertex is due to 
    minimal (non-minimal) coupling,
    which is independent of (quadratic in) the photon momentum.
    The diagrams (a,b,c) give contributions to
    the point-like part of the charge radius. 
    Diagrams (d,e,f) enter with the finite-size contribution of the
    core (e) and the nucleon (f), and through a short-range
    contribution (d). 
      \label{fig:haloformfactordiagram}}
  \end{center}
\end{figure}

%
The sizes of the halo system and its components are manifest not
only in the ERE parameters but also in the charge form factor.
The charge form factor 
is obtained as the matrix element of
the zeroth component of the electromagnetic four-current, $J^\mu$,
according to 
\begin{equation} 
F_\mrm{ch}(\QQ)=\frac{1}{e Z_\mrm{h}}\langle J^0\rangle~
=1-\frac{r_\mrm{ch}^2}{6}\QQ^2+\dots~,
\label{chargeFFdef}
\end{equation}
where $Z_\mrm{h}$ is the proton number of the halo nucleus
and $\QQ$ is the momentum transfer.
A measure of the size of the nucleus is
the charge radius $r_\mrm{ch}$.
We now look at the power counting for the observable charge form factor
of one-nucleon halo systems. 
The discussion here follows {R}ef.~\cite{Hammer:2011ye}, but makes explicit
the factors of $1/\Ac$ that were not incorporated into the power counting there.
We discuss contributions to the charge form factor in the 
point-like limit, due to the finite-size of the constituents, 
and from additional two-body (short-range) contributions, as displayed in 
Fig. \ref{fig:haloformfactordiagram}.
More details and explicit examples will be presented in Sec.~\ref{sec:3}.

We start by discussing the point-like part, $r_\mrm{pt}$, which 
comes from the photon coupling to the charge of the constituents.
The corresponding operators in the Lagrangian have the general form
$\psi^\dagger A_0\psi$ where $\psi$ denotes either the core or the nucleon field 
and $A_0$ is the zeroth component of the photon four-vector field. 
The point-like contribution is kinematically generated by nucleon-core
one-loop diagrams, where the photon couples to the core (in the neutron halo
case) or to both the core and the proton (for the proton halo) --- see 
Fig. \ref{fig:haloformfactordiagram}(b,c).
For a neutron halo, momentum dependence requires recoil of the core,
and thus the charge radius, which involves two powers of momenta,
is proportional to $f^2$, where $f$ is defined in Eq. \eqref{cNtotmass}.
Indeed, the loop for neutron-halo systems
was calculated by Hammer and Phillips \cite{Hammer:2011ye} with the 
leading-order (LO) results:
\begin{equation}
r_\mrm{pt,LO}^2=\left\{\begin{array}{lc} \frac{f^2}{2 \gamma_0^2}~, 
&\mrm{~S\text{-}wave~neutron~halo,}\\
-\frac{5f^2}{2\gamma_1(3\gamma_1+r_1)}~, 
&\mrm{~P\text{-}wave~neutron~halo.} 
\end{array}\right.
\label{eq:pointradii}
\end{equation}
The scalings are then given by 
\begin{equation}
r_\mrm{pt}^2\sim \left\{\begin{array}{lc} 1/(A_\mrm{c}^2k_\mrm{lo}^2)~, 
&\mrm{~S\text{-}wave~neutron~halo,}
\\
1/(A_\mrm{c}^2k_\mrm{lo}k_\mrm{hi})~, 
&\mrm{~P\text{-}wave~neutron~halo,}\end{array}\right.
\label{pointcontestimate}
\end{equation} 
assuming Eqs. \eqref{gamma0}, \eqref{r1magnitude}, and \eqref{gamma1}. 
However, it should be noted that typically $r_1 < 0$ so, 
if $|r_1|$ is close to $3\gamma_1$,
then $|r_1 + 3 \gamma_1|$ can in practice be $\sim k_\mrm{lo}$ rather than the 
formally correct assignment $\sim k_\mrm{hi}$ we have used here. 

The point-like contribution to the charge radius is 
interesting because it can be calculated from known properties
of the constituents, but it exists against a backdrop of 
other, less well-known contributions.
One type of these other contributions comes
from the finite sizes of the constituent core and nucleon,
which enter through the same loop diagrams as the point contributions, 
see Figs.~\ref{fig:haloformfactordiagram}(e,f).
The finite-size contributions correspond to
operators of the form $\psi^\dagger (\nabla^2A_0)\psi$.
The two extra powers of the small momentum compared to the point-like
terms coming from minimal substitution means that,
by naive dimensional analysis, this operator carries a
suppression of $k_\mrm{hi}^{-2}$.
Such terms produce a direct contribution to the charge radius 
that is not suppressed by $\Ac^{-2}$. The nucleon finite-size contribution to 
$r^2_\mrm{ch}$ will be denoted by $\rho_\mrm{N}^2$ and should be proportional to
$\rho_\mrm{N}^2\sim  R_\mrm{N}^2$, i.e., to the 
proton or neutron charge radius squared, 
respectively, $\rho_\mrm{p}^2=0.766~\mrm{fm}^2$~\cite{mohr_2015_22826}
and $\rho_\mrm{n}^2=-0.116~\mrm{fm}^2$~\cite{Olive:2016xmw}.
This contribution scales as
\begin{equation} 
\rho_\mrm{N}^2/Z_\mrm{h}\sim  R_\mrm{N}^2/Z_\mrm{h},
\mrm{~S\text{-}wave~or~P\text{-}wave~halo}.
\label{eq:scalenucleon}
\end{equation} 
Meanwhile, the core-size contribution to the charge radius squared, 
$\rho_\mrm{c}^2$, will scale as
\begin{equation}
\rho_\mrm{c}^2\sim  R_\mrm{c}^2\sim 1/k_\mrm{hi}^2, 
\mrm{~S\text{-}wave~or~P\text{-}wave~halo}.
\label{eq:coresizecont}
\end{equation}
Taking the ratio of Eqs. \eqref{eq:scalenucleon} and \eqref{eq:coresizecont} 
shows that in general the nucleon-finite-size contribution to the charge 
radius squared is smaller than 
the core-finite-size contribution 
by both $(R_\mrm{N}/R_\mrm{c})^2$ and a factor of the total charge $Z_{\mrm{h}}$ of the 
system. For the canonical estimate $R_\mrm{c} \sim \Ac^{1/3} R_\mrm{N}$ we have 
\begin{equation}
\rho_\mrm{N}^2/(Z_\mrm{h}\rho_\mrm{c}^2) \sim 1/(Z_\mrm{h} \Ac^{2/3}).
\end{equation}

There is another contribution to the charge radius, but this time it is not 
determined by data from other processes. Short-range contributions to the
charge density are encoded in a contact operator of the form 
$\Psi^\dagger(\nabla^2A_0)\Psi$,
where $\Psi$ denotes the dicluster field for either the S- or the P-wave
system --- see Fig. \ref{fig:haloformfactordiagram}(d). 
Because of the two derivatives,
this operator is suppressed by a factor of $k_\mrm{hi}^{-2}$
with respect to $\Psi^\dagger A_0\Psi$, which originates
in the minimal substitution of the dicluster kinetic term
--- Fig. \ref{fig:haloformfactordiagram}(a).
The minimal substitution term ensures that the charge comes out correct;
the term with two additional derivatives comes with
short-range parameters, 
which we denote in S- and P-wave halos by,
respectively, $\rho_{\sigma,\pi}^2 \sim\khi^{-2}$.
In the case of an S-wave halo, the dicluster kinetic term is 
itself a relative $k_\mrm{hi}^{-1}$ effect
(it gives rise to the range, Eq. \eqref{a0r0magnitude}),
for an overall $k_\mrm{hi}^{-3}$ suppression.
For a P-wave system, the dicluster kinetic term 
leads to the ``range'' which scales with $k_\mrm{hi}$,
see Eq. \eqref{r1magnitude}, and there is no extra 
suppression~\cite{Hammer:2011ye}.
We expect short-range contributions to the charge radius that scale as
\begin{equation}
\left\{\begin{array}{ll}
\gamma_0r_0\rho_{\sigma}^2\sim k_\mrm{lo}/k_\mrm{hi}^3, & \mrm{~S\text{-}wave~halo},
\\
\rho_{\pi}^2\sim 1/k_\mrm{hi}^2, & \mrm{~P\text{-}wave~halo},
\end{array}\right.
\label{eq:naiveshortrangescaling}
\end{equation}
again shown explicitly in Sec.~\ref{sec:3}.
Thus, for S-wave one-nucleon halos this short-range operator enters one
order after the core-size contribution.
For P-wave halos both contribute at the same order.

In summary, these power-counting arguments make explicit that the
point-like contribution for one-nucleon halos involves a kinematical
suppression factor $1/\Ac^2$ for neutron halos. 
But this has the consequence that, for P-wave one-neutron halos, a
short-range operator enters at the same order as the finite-size
core contribution. The existence of such additional short-range
operators will have negative influence on the predictive power of LO
calculations.  

\section{The charge radius formula}
\label{sec:3}
%
In this section we will derive charge-radius formulas in HEFT with the
heavy core power counting. In the process we will critique some results
from {R}ef.~\cite{Hammer:2011ye} where finite-size effects were 
not treated explicitly.
For example, the charge radius formula used in {R}ef.~\cite{Hammer:2011ye} for a
one-neutron halo system is
\begin{equation}
r_\mrm{ch}^2=r_\mrm{pt}^2+\rho_\mrm{c}^2 ,
\label{eq:chargeradiusformulaHP}
\end{equation} 
where $\rho_\mrm{c}^2$ is the charge radius squared of the core. 
In principle, one should also add the neutron charge
radius contribution $\rho_\mrm{n}^2/Z_\mrm{c}$, where $Z_\mrm{c}$ is
the core charge, but this term is usually neglected since
$|\rho_\mrm{n}^2|$ is tiny. 
The key point is that
Eq.~\eqref{eq:chargeradiusformulaHP} has not been
derived within the field theory: finite-size contributions
were instead added  to the point-like result in a rather \emph{ad hoc} fashion.
In what follows we will show that $\rho_\mrm{c}^2$ (and, for that matter, 
$\rho_{\mrm{n}}^2/Z_\mrm{c}$) indeed add to the charge radius squared,
but in principle other contributions of similar size can appear. 
We also argue that it is important to keep track of 
the large suppression in neutron 
halos of the point-like radius for $\Ac\gg1$, when the main contribution to the
charge radius of the system comes from the finite size of the
constituents.

This derivation, carried out in the next subsections,
starts from the HEFT Lagrangian. 
We will consider explicitly only dominant S- or P-wave interactions,
and a spin-0 core --- generalizations are straightforward
but cumbersome to write.
The Lagrangian for a spin-1/2 nucleon $N_s$, where $s=-1/2,1/2$, and a
spin-0 core $c$ with S- and P-wave interactions is given by
\begin{align}
 \mathcal{L} =&
N_s^\dagger\left[i\partial_0-
\frac{e}{2}(1+\tau_3) A_0+\frac{\nabla^2}{2m_\mrm{N}}
-\frac{e}{12}\left[\left(\rho_\mrm{p}^2+\rho_\mrm{n}^2\right)+
\left(\rho_\mrm{p}^2-\rho_\mrm{n}^2\right)\tau_3 \right]
(\nabla^2A_0)+\dots\right]N_s
\nonumber \\
&+c^\dagger\left[i\partial_0-eZ_\mrm{c}A_0
+\frac{\nabla^2}{2m_\mrm{c}}-\frac{eZ_\mrm{c}\rho_\mrm{c}^2}{6}(\nabla^2A_0)
+\dots\right]c
\nonumber\\
&+\sigma_s^\dagger\left[\Delta_0+\eta_0\left(i\partial_0-e Z_\mrm{h} A_0
+\frac{\nabla^2}{2M_\mrm{cN}}-\frac{eZ_\mrm{h}}{6}\rho_\sigma^2\,(\nabla^2A_0)
\right) +\dots\right]\sigma_s
\nonumber \\
&-g_0\left(\sigma_s^\dagger c N_s+\mrm{H.c.}\right)+\dots~
\nonumber \\
&+\pi_s^\dagger\left[\Delta_1+\eta_1\left(i\partial_0-eZ_\mrm{h} A_0
+\frac{\nabla^2}{2M_\mrm{cN}}-\frac{eZ_\mrm{h}}{6}\rho_\pi^2\,(\nabla^2A_0)
\right)+\dots\right]\pi_s
\nonumber \\
&-g_1\left[\mathcal{C}_{si}^{s'}\pi_{s'}^\dagger c
\left(i(1-f)\overrightarrow{\nabla}_i-if\overleftarrow{\nabla}_i\right) N_s
+\mrm{H.c.}\right]+\dots~.
\label{eqLag}
\end{align}
The field $\sigma_s$ is the spin-1/2 dicluster
field, which we introduce for convenience. Its kinetic
term has a sign $\eta_0$ and it has a residual mass $\Delta_0$, which
is a parameter to be fixed. 
The most important S-wave coupling of nucleon and core
has strength $g_0$.
For the P-wave interaction, corresponding to the last two terms, we
have, for simplicity, included only the $J=1/2$ channel, 
through a spin-1/2 dicluster field
$\pi_s$ and the Clebsch-Gordan coefficient 
$\mathcal{C}_{si}^{s'}=\left(\frac{1}{2}s 1i \big|\left(\frac{1}{2}1\right) 
\frac{1}{2}s' \right)$~\cite{Hammer:2011ye}. 
The indices take values according to $s,s'=-1/2,1/2$ and $i=-1,0,1$.
As above, $\eta_1$ is a sign, and $\Delta_1$ and $g_1$
are parameters to be fixed, while
$f$ was defined in Eq. \eqref{cNtotmass}. 
The field $A_0$ is the zeroth component of the photon four-vector field.
Here $\tau_3$ is the third isospin Pauli matrix, and
we have defined the charge number of the core $Z_\mrm{c}$.
Note that 
$Z_\mrm{h}$ is the charge ($Z_\mrm{c}$ or $Z_\mrm{c}+1$)
of the nucleon-core system.
The charge radius of the
nucleon (core) field is $\rho_\mrm{N}$ ($\rho_\mrm{c}$),
while $\rho_\sigma$ and $\rho_\pi$ are additional short-range
parameters with sizes given by Eq. \eqref{eq:naiveshortrangescaling},
which will be discussed below.

This Lagrangian includes all operators that will contribute to the
charge radius of the halo system up to NLO.
Higher-order terms, such as the one that leads
to the shape parameter in the ERE and terms that do not
contribute to the charge form factor, are denoted by the ellipsis. 
The coefficients of the $\nabla^2A_0$ terms in Eq.~\eqref{eqLag} encode  
the finite size and composite nature of the core. 
We assign the same scaling to them as
Hammer and Phillips, who used naive dimensional analysis to argue that 
$\rho_\mrm{c}^2, \rho_\sigma^2, \rho_\pi^2$ are all $\sim 
R_\mrm{c}^2$~\footnote{For $\rho_\pi^2$, which undergoes renormalization, 
this estimate is
only valid for a renormalization scale of order $k_\mrm{hi}$, see below.}.
However, 
whereas {R}ef.~\cite{Hammer:2011ye} argued that this scaling rendered these 
effects of higher order, here 
we contend that they can provide the main contribution to the charge radius 
for heavy core systems.

\begin{figure}
  \begin{center}
    \includegraphics[scale=0.5]{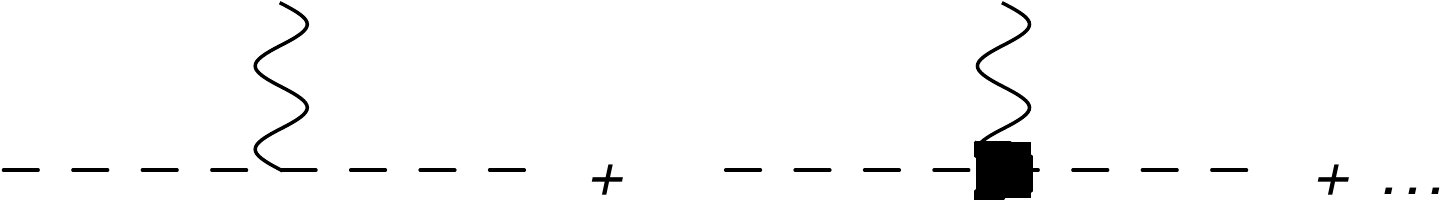}
    \caption{Diagrams for the charge form factor of the core. The
      first diagram has a vertex $ieZ_\mrm{c}$ and the second diagram
      gives the charge radius of the core, through the vertex
      $-ieZ_\mrm{c}\rho_\mrm{c}^2Q^2/6$.  \label{fig:coreformfactordiagram}} 
  \end{center}
\end{figure}

The finite constituent sizes are reflected in the non-trivial momentum-transfer
dependence of the core and nucleon form factors.
The form factor of the core is given by the part of the Lagrangian 
where a photon couples to the core field, that is,
terms of the form $c^\dagger A_0 c$ (and derivatives of $A_0$). 
The resulting charge form factor of the core is thus given by the diagrams 
shown in Fig.~\ref{fig:coreformfactordiagram} as 
\begin{align}
F_\mrm{ch,c}(\QQ)=&\frac{1}{eZ_\mrm{c}}\langle J^0_\mrm{c}\rangle
=1-\frac{\rho_\mrm{c}^2}{6}\QQ^2+\dots,
\label{coreFF}
\end{align} 
and as such the charge radius of the core is given by $\rho_\mrm{c}$.
Note that since Fig.~\ref{fig:coreformfactordiagram} contains no loops 
the Lagrangian parameter $\rho_\mrm{c}^2$ appears directly here without
being affected by renormalization---at this order. 
The nucleon charge form factors and charge radii
can be considered in a similar fashion, 
with the exception that the electric charge of the neutron is zero:
\begin{equation}
F_\mrm{ch,N}(\QQ)=\frac{1}{e}\langle J^0_\mrm{N}\rangle
=Z_\mrm{N}-\frac{\rho_\mrm{N}^2}{6}\QQ^2+\dots~,
\label{nucleonFF}
\end{equation}
where $Z_\mrm{N}=0$ $(1)$ for the neutron (proton).
We now examine the effect of these terms in the charge radii
of S- and P-wave neutron halos, while the case of proton halos is
discussed in the Appendix.

\subsection{S-wave neutron halos
\label{sec:Swavenhalo}}
%
Here we compute the expectation value of the zeroth component of the 
electromagnetic
current, $\langle J^0\rangle$, which appears in Eq. \eqref{chargeFFdef}, 
for an S-wave
one-neutron halo. The long-distance
contributions to this quantity are well known, cf. 
Refs.~\cite{Chen:1999tn,Beane:2000fx,Hammer:2011ye}, 
where the diagrams in Fig.~\ref{fig:haloformfactordiagram}(a,b) were 
evaluated (although only
for $A_{\mrm{c}}=1$ in Refs.~\cite{Chen:1999tn,Beane:2000fx}). 
Here we include diagrams Fig.~\ref{fig:haloformfactordiagram}(d,e,f)
as well, and so account for finite-size effects and the leading 
short-range, two-body operator.

The charge form factor of an S-wave one-neutron halo can be computed
from the amputated correlator of the $\sigma_s$ field with one insertion of 
all possible couplings to an $A_0$ photon. 
The diagrams that contribute up to ${\cal O}(\left(k_{\rm lo}/k_{\rm hi}\right)^3)$
are shown in Fig.~\ref{fig:haloformfactordiagram}.
We must also apply a wavefunction renormalization factor 
$\mathcal{Z}_0$, which defines the overlap of the field $\sigma_s$ with
the physical one-neutron halo state. The contributions from tree and loop
diagrams can then be separated, viz.
\begin{equation}
F_\mrm{ch}(\QQ)=\frac{1}{eZ_\mrm{c}}
\left[\Gamma_\mrm{tree}(\QQ)+\Gamma_\mrm{loop}(\QQ)\right]~.
\label{FFastreeplusloop}
\end{equation}
The wavefunction renormalization factor is~\cite{Beane:2000fx,Hammer:2011ye}:
\begin{equation}
\mathcal{Z}_0= 
\frac{2\pi\gamma_0}{g_0^2\mR^2} 
(1-\gamma_0r_0)^{-1}\;,\;\rm{up \; to \; NLO}~,
\label{eq:LSZneutronS}
\end{equation}
where 
we kept some higher-order terms as well by not expanding
the $(1-\gamma_0r_0)^{-1}$ ratio.
Note that $\mathcal{Z}_0$ is finite to this order.

At $\QQ=0$ finite-size effects cannot play a role as the photon only
``sees" the entire charge of the system. The leading 
contribution to the form factor
at $\QQ=0$ is then from
the loop diagram in Fig.~\ref{fig:haloformfactordiagram}(b),
where the virtual photon is attached to the core via its charge. 
This diagram also gives rise to subleading corrections
to $F_\mrm{ch}(\QQ)$: it generates the point contribution
to the form factor, but away from $\QQ=0$, this is
suppressed by $1/\Ac^2$ and not as large as other
effects once $\Ac \gg 1$.

The most important such other effect is due to the loop diagram 
Fig.\ref{fig:haloformfactordiagram}(e), i.e., the coupling of the $A_0$
photon to the finite size of the core inside the loop. The contribution
of this graph can be combined with the corresponding coupling
for the neutron,  Fig.~\ref{fig:haloformfactordiagram}(f), to obtain
the contribution stemming from the
constituent 
form factors, Eqs. \eqref{coreFF} and \eqref{nucleonFF}. The result
can be expressed as a coordinate-space integral,
\begin{align}
\Gamma_\mrm{loop}(\QQ)=&eZ_\mrm{c}
\frac{g_0^2\mR^2}{2\pi}
\mathcal{Z}_0
\int\!\d r\d(\cos{\theta})\exp{\left(-2\gamma_0r\right)}
\nonumber\\
&\qquad\quad
\times\bigg[\left(1-\frac{\rho_\mrm{c}^2}{6}\QQ^2\right)
\exp{(if\QQ\cdot\rr)}
-\frac{\rho_\mrm{n}^2}{6Z_\mrm{c}}\QQ^2\exp{\left(i(1-f)\QQ\cdot\rr\right)}\bigg]
~.
\label{eq:GammaLoopNeutronSfull}
\end{align}
We expand the integral \eqref{eq:GammaLoopNeutronSfull}
up to order $\QQ^2$ to arrive at
\begin{equation}
\Gamma_\mrm{loop}(\QQ)=eZ_\mrm{c}
\frac{g_0^2\mR^2}{2\pi\gamma_0}
\mathcal{Z}_0
\left[1-\left(\rho_\mrm{c}^2+\frac{\rho_\mrm{n}^2}{Z_\mrm{c}}
+\frac{f^2}{2\gamma_0^2}\right)\frac{\QQ^2}{6}+\dots\right]~.
\label{eq:GammaLoopNeutronSfullexpanded}
\end{equation}

At ${\cal O}(k_\mrm{lo}/k_\mrm{hi})$, $F_\mrm{ch}(\QQ)$  also receives a 
contribution from the tree-level 
diagram, Fig.~\ref{fig:haloformfactordiagram}(a).
Considering also the $O(\left(k_\mrm{lo}/k_\mrm{hi}\right)^3)$ tree diagram in 
Fig.~\ref{fig:haloformfactordiagram}(d),
which represents the short-range contribution to the halo 
charge radius, we arrive at
\begin{equation}
\Gamma_\mrm{tree}(\QQ)=-e Z_\mrm{c} \frac{g_0^2 m_\mrm{R}^2}{2 \pi r_0} 
\mathcal{Z}_0
\left(1-\frac{\rho_\sigma^2}{6}\QQ^2\right)~.
\label{eq:gammatreeneutronSfull}
\end{equation} 
The first term here is a consequence of charge conservation and ensures 
that as $\QQ \rightarrow 0$ we have $F_\mrm{ch}(0)=1$, that is, the charge 
form factor is correctly normalized at
zero momentum transfer. Moving away from $\QQ \rightarrow 0$ we insert
Eqs. \eqref{eq:gammatreeneutronSfull} and 
\eqref{eq:GammaLoopNeutronSfullexpanded} in Eq. \eqref{FFastreeplusloop}, 
and compare with the term quadratic in momentum in Eq. \eqref{chargeFFdef}, 
to obtain the charge-radius formula for S-wave neutron halos,
\begin{equation}
r_\mrm{ch}^2=\frac{1}{1-\gamma_0r_0}
\left(r_\mrm{pt,LO}^2+\rho_\mrm{c}^2+\frac{\rho_\mrm{n}^2}{Z_\mrm{c}}
-\gamma_0r_0\rho_\sigma^2\right) +\ldots~,
\label{eq:chargeradiusformulaneutronSwave}
\end{equation}
where the ``$\dots$'' represent higher-order contributions.
Note that at this order the Lagrangian parameter $\rho_\sigma^2$ appears 
directly here, without renormalization. 
This is because the LO loop that gives the point-charge contribution is finite. 
The point-charge contribution to $r_\mrm{ch}^2$ was computed in 
{R}ef.~\cite{Hammer:2011ye} as $(1-\gamma_0r_0)^{-1}r_\mrm{pt,LO}^2$, with
$r_\mrm{pt, LO}^2$ given by Eq.~\eqref{eq:pointradii}.

Expanding in $\gamma_0r_0$,
\begin{equation}
r_\mrm{ch}^2=\rho_\mrm{c}^2 
+\gamma_0r_0\left(\rho_\mrm{c}^2-\rho_\sigma^2\right)+\ldots
+\left(\frac{\rho_\mrm{n}^2}{Z_\mrm{c}}+r_\mrm{pt,LO}^2\right) 
\left(1+\gamma_0r_0 +\ldots\right),
\label{eq:Sradiusexpanded}
\end{equation}
where the orders of various contributions are summarized in 
Table \ref{ordercontributionsSwaveradius}, assuming that 
$R_{\mrm{N}}/R_{\mrm{c}} \lesssim A_{\mrm{c}}^{-1/3}$.
If $A_\mrm{c}\sim 1$, the most important terms are given by the
point-radius $r_\mrm{pt,LO}^2$. 
For larger $A_\mrm{c}$, these terms rapidly lose importance,
as do contributions from the neutron size.
Unless we are dealing with light halos, we expect
the dominant contribution to the difference in charge radii
between halo and core to be given by 
$\gamma_0r_0(\rho_\mrm{c}^2-\rho_\sigma^2)$.
This is an example of a term that is missed if one simply adds the
core radius by hand, rather than including it in the EFT as any
other operator.

\begin{table}
\begin{center}
  \caption{Orders of the various contributions 
    to the charge radius of S-wave neutron halos
    listed in Eq. \eqref{eq:Sradiusexpanded}. 
    In each column effects of a particular order in the usual HEFT
    expansion parameter $\klo/\khi$ appear.  
    Meanwhile the rows organize contributions due to
    additional small factors: inverse powers
    of the number of core nucleons ($A_\mrm{c}$) and protons ($Z_\mrm{c}$).}
\label{ordercontributionsSwaveradius}
\vspace{0.3cm}
{\renewcommand{\arraystretch}{1.5}
\begin{tabular}
{c | c@{\hspace{5mm}} ccc}
\hline\hline 
         & ${\cal O}(k_\mrm{lo}^{-2})$
         & ${\cal O}((k_\mrm{lo}k_\mrm{hi})^{-1})$ 
         & ${\cal O}(k_\mrm{hi}^{-2})$
         & ${\cal O}(k_\mrm{lo}k_\mrm{hi}^{-3})$ 
         \\
\hline
 ${\cal O}(1)$ 
& --- 
& --- 
& $\rho_\mrm{c}^2$ 
& $\gamma_0r_0(\rho_\mrm{c}^2-\rho_\sigma^2)$ 
\\ 
 ${\cal O}(A_\mrm{c}^{-3/2}Z_\mrm{c}^{-1})$  
& --- 
& --- 
& $\rho_\mrm{n}^2/Z_\mrm{c}$
& $\gamma_0r_0\, \rho_\mrm{n}^2/Z_\mrm{c}$
\\ 
 ${\cal O}(A_\mrm{c}^{-2})$  
& $r_\mrm{pt,LO}^2$  
& $\gamma_0r_0\, r_\mrm{pt,LO}^2$  
& $\ldots$
& $\ldots$ 
\\ 
\hline\hline
\end{tabular}}
\end{center}
\end{table}

Unfortunately, while $\rho_\mrm{c}$ can be extracted
from the core form factor \eqref{coreFF},
$\rho_\sigma$ is a short-range term that
cannot easily be extracted from a quantity other
than the halo form factor itself. 
Since this is a short-range effect it is possible that it can be efficiently 
extracted from {\it ab initio} calculations of the charge radius. 
In such a calculation the difference between $\rho_\sigma$
and $\rho_\mrm{c}$ can be viewed as originating in
two effects:
\begin{enumerate}
\item A change in the size of the core when it is placed in the bound 
state with the neutron.
\item Pieces of the {\it ab initio} wave function not in the core + neutron 
piece of the 
Hilbert space, e.g., those due to excited states of the core.
\end{enumerate}
These effects cannot, however, be separated in a model-independent way,
and only their combination enters through $\rho_\mrm{c}^2-\rho_\sigma^2$.

As a concrete example we consider the S-wave ground state 
of the one-neutron halo $\nuc{11}{Be}$,
whose form factor and photodisintegration were
investigated in {R}ef.~\cite{Hammer:2011ye}.
The neutron separation
energy is $B_{\mrm{s}0}=0.502~\mrm{MeV}$ \cite{AjzenbergSelove:1990zh}, 
corresponding
to $\klo\sim \gamma_0
\simeq 30~\mrm{MeV}$ 
through Eq. \eqref{gamma0}.
Using the charge radius of the $\nuc{10}{Be}$ core,
$\rho_\mrm{c}=2.357(18)~\mrm{fm}$~\cite{Nortershauser:2009bd}, 
as an estimate for its size,
the breakdown scale is 
$\khi\sim 1/R_\mrm{c}\simeq 80~\mrm{MeV}$.
This is also the momentum 
$\sqrt{2\mR E_\mrm{ex}}\simeq 80~\mrm{MeV}\sim \khi$
corresponding
to the first excitation of the core at $E_\mrm{ex}=3.368~\mrm{MeV}$
\cite{TILLEY2004155}, 
so there is no need to include a field for this state.
These scales then give us the expansion parameter
$k_\mrm{lo}/k_\mrm{hi}\approx 0.4$.
This means that $r_\mrm{pt,LO}^2$ is numerically of the same size as 
$\mathcal{O}((k_{\rm lo}/k_{\rm hi})^3)$ corrections. 
Since $Z_\mrm{c}\sim A_\mrm{c} k_\mrm{lo}/k_\mrm{hi}$,
the neutron-radius contributions are suppressed by more
than five powers of $k_\mrm{lo}/k_\mrm{hi}$ compared to $\rho_\mrm{c}$.
At LO there is a charge-radius prediction, but it is trivial since it is just 
the charge radius of the $\nuc{10}{Be}$ core,
$\rho_\mrm{c}^2\simeq 5.56(4)~\mrm{fm}^2$.
This does, though, explain most of the measured value of
$r_\mrm{\nuc{11}{Be}}^2 \simeq 6.07(8)~\mrm{fm}^2$
($r_\mrm{\nuc{11}{Be}}=2.463(16)~\mrm{fm}$~\cite{Nortershauser:2009bd}).
Estimating
$\gamma_0r_0$ from the EFT expansion parameter $\sim 0.4$, we find
that the point-charge contribution to $r_\mrm{ch}^2$,
i.e., the first term of
Eq.~\eqref{eq:chargeradiusformulaneutronSwave},
is $r_\mrm{pt,LO}^2 / \left( 1-\gamma_0r_0 \right) \simeq
0.3~\mrm{fm}^2$. This explains more than half of the difference
$r_\mrm{\nuc{11}{Be}}^2 - r_\mrm{\nuc{10}{Be}}^2$. 
The rest must come from the short-distance effect $\rho_\sigma^2$:
the experimental value for $r_\mrm{\nuc{11}{Be}}$ can be reproduced if
the short-range
parameter is given by $\rho_\sigma^2\approx 5.1~\mrm{fm}^2$, which is of the expected order of magnitude, $1/k_\mrm{hi}\sim 2.5~\mrm{fm}$.
This supports the power counting presented here. We thus
see that $\rho_\sigma^2$ must be a little smaller than $\rho_\mrm{c}^2$
in order to explain the data, although the errors on the atomic measurements
of the $\nuc{10}{Be}$ and $\nuc{11}{Be}$ radii make it difficult to extract
a precise value for $\rho_\mrm{c}^2 - \rho_\sigma^2$.

\subsection{P-wave neutron halos}
%
An important aspect of the S-wave halo system is that all the charge 
form-factor diagrams we considered are finite. For P waves this is not the case.
The increased singularity of the P-wave interaction
can be seen already in the need for the effective-range term 
\eqref {r1magnitude}
at LO to allow proper renormalization of nucleon-core scattering. 
As before we will consider 
operators up to
second order in the photon momentum $\QQ$ and 
will show that, if the
charge-radius contributions of the constituents are to be considered 
explicitly,
we need an additional short-range operator to renormalize 
the halo charge radius.

Since the cancelation of divergences will be critical to what 
follows we recapitulate the formulas for neutron-core
scattering derived in Refs.~\cite{Bertulani:2002sz,Hammer:2011ye}.
The power counting discussed in Sec. \ref{Nucleoncorescattering}
indicates that neutron-core scattering proceeds through
the bare dicluster propagator at LO, and by an insertion
of one nucleon-core bubble at NLO.
Up to this order,
elastic scattering with a P-wave interaction gives the matching
\begin{eqnarray}
\frac{1}{a_1}&=&
\frac{6\pi\Delta_1}{g_1^2\mR} + \frac{2L_3}{\pi}~,
\label{eq:effectiverangePa1}\\ 
r_1&=&-\frac{6\pi\eta_1}{g_1^2\mR^2} - \frac{4L_1}{\pi}~,
\label{eq:effectiverangePr1}
\end{eqnarray}
where the $L_n=\int\d p \, p^{n-1}$ are divergent integrals in
the irreducible dicluster self-energy,
\begin{equation}
\Sigma_1(E)= \frac{g_1^2\mR}{6\pi}
\left[\frac{2L_3}{\pi}+\frac{4L_1}{\pi}\mR E+i(2\mR E)^{3/2}\right]~.
\end{equation} 
It is evident from
Eqs.~\eqref{eq:effectiverangePa1} and \eqref{eq:effectiverangePr1}
that two parameters, $\Delta_1$ and $g_1$, are needed to renormalize
scattering up to NLO.
The P-wave wavefunction renormalization 
is given by
\begin{equation}
\mathcal{Z}_1=-\frac{6\pi}{g_1^2\mR^2 r_1}
\left(1+\frac{3\gamma_1}{r_1}\right)^{-1}
\;,\;\rm{up \; to \; NLO}~.
\label{eq:Z1}
\end{equation}
Note that, contrary to the S-wave  wavefunction renormalization
\eqref{eq:LSZneutronS}, $\mathcal{Z}_1$ is not finite to this order.

The P-wave charge form-factor diagrams are similar to 
those for the S-wave interaction, 
Fig.~\ref{fig:haloformfactordiagram}.
The tree diagrams amount to
\begin{equation}
\Gamma_\mrm{tree}(\QQ)=\eta_1eZ_\mrm{c}\mathcal{Z}_1
\left(1-\frac{\rho_\pi^2}{6}\QQ^2\right)~,
\label{eq:GammaTreePneutron}
\end{equation}
while the one-loop diagrams give
\begin{align}
\Gamma_\mrm{loop}(\QQ)=&eZ_\mrm{c}\frac{g_1^2\mR^2\gamma_1^2}{6 \pi}\mathcal{Z}_1
\int\!\d r\d(\cos{\theta})\left(1+\frac{1}{\gamma_1r}\right)^2
\exp{(-2\gamma_1r)}
\nonumber\\
&\times\bigg[\left(1-\frac{\rho_\mrm{c}^2}{6}\QQ^2\right)\exp{(if\QQ\cdot\rr)}
-\frac{\rho_\mrm{n}^2}{6Z_\mrm{c}}\QQ^2\exp{(i(1-f)\QQ\cdot\rr)}\bigg]~.
\label{eq:GammaLoopNeutronPdiv}
\end{align}
Expanding in powers of the momentum transfer $\QQ$,
\begin{equation}
\Gamma_\mrm{loop}(\QQ)=
-eZ_\mrm{c}\frac{g_1^2\mR^2\gamma_1}{2\pi} \mathcal{Z}_1
\left\{\left(1-\frac{4 L_1}{3\pi\gamma_1}\right)
\left[1-\left(\rho_\mrm{c}^2+\frac{\rho_\mrm{n}^2}{Z_\mrm{c}}\right)
\frac{\QQ^2}{6}\right]
+\frac{5f^2}{6\gamma_1^2}\frac{\QQ^2}{6}
+\dots\right\}~.
\label{eq:formfactorPneutronStuffprime}
\end{equation}
The only difference with respect to the S wave, apart from $\mathcal{Z}_1$,
is that the P-wave bound-state
wavefunction is $\left[1+1/(\gamma_1r)\right]\exp{(-\gamma_1r)}$, which
is irregular at the origin.
As a consequence, the integral that appears already
in the momentum-independent term is divergent, and
related to one of the divergent integrals in $\Sigma(E)$,
the $L_1$ of Eq. \eqref{eq:effectiverangePr1}.

The divergence in the momentum-independent contribution
cancels between 
Eqs. \eqref{eq:GammaTreePneutron} and  \eqref{eq:formfactorPneutronStuffprime},
and we obtain a properly normalized form factor, $F_\mrm{ch}(0)=1$
\cite{Hammer:2011ye}. 
The terms quadratic in momentum give the charge radius
\begin{equation}
r_\mrm{ch}^2=r_\mrm{pt,LO}^2+ \rho_\mrm{c}^2+\frac{\rho_\mrm{n}^2}{Z_\mrm{c}}
+ \bar{\rho}_\pi^2 +\ldots,
\label{eq:chargeradiusformulaneutronPwave}
\end{equation}
where the LO point-charge contribution, defined in Eq.~(\ref{eq:pointradii}), 
agrees with {R}ef.~\cite{Hammer:2011ye},
and the (finite) short-range contribution $\bar{\rho}_\pi^2$ is related to the
counterterm $\rho_\pi^2$ by
\begin{equation}
\bar{\rho}_\pi^2=\frac{1}{r_1 + 3 \gamma_1} \left(r_1 + \frac{4 L_1}{\pi}\right) 
\left(\rho_\pi^2 - \rho_{\mrm c}^2 - \frac{\rho_{\mrm n}^2}{Z_{\mrm c}}\right).
\label{eq:rhobarpisq}
\end{equation}
An interesting point here is that the finite contribution of $\rho_\mrm{c}^2$ to
$r_\mrm{ch}^2$ from Eq. \eqref{eq:formfactorPneutronStuffprime} is suppressed by
an additional factor $\gamma_1/r_1\sim k_{\rm lo}/k_{\rm hi}$ with respect to the 
estimate \eqref{eq:coresizecont}. The appearance of the full $\rho_\mrm{c}^2$ 
in \eqref{eq:chargeradiusformulaneutronPwave} is a consequence of the 
particular renormalization condition \eqref{eq:rhobarpisq}.

In this renormalization scheme the effect beyond the ``standard" charge-radius
formula depends on the extent to which the 
dicluster counterterm 
differs from the core radius.
In contrast to the S-wave case, here the difference
$\rho_\pi^2 - \rho_{\mrm c}^2 -\rho_{\mrm n}^2 / Z_{\mrm c}$
must go to zero as the regulator is taken to infinity, in order to yield a
finite $\bar{\rho}_\pi^2$.
It is important to consider what would happen if we were to include
the finite-size contributions, but not the $\rho_\pi^2$ short-range
operator. In Eq.~(\ref{eq:formfactorPneutronStuffprime}) we see that the
constituent charge radii enter with a prefactor
that corresponds to a divergent
integral. Since the parameters $\rho_\mrm{c}^2$ and $\rho_\mrm{n}^2$
are observables---these are the charge radii of the core and the
neutron---they cannot absorb this divergence. The only parameter
available for this purpose is the $\rho_\pi^2$. As such it is not
possible to add the finite-size contributions without also including
the short-range operator. Formally, the scaling of $\rho_\pi^2$ is 
$\rho_\pi^2 \sim \bar{\rho}_\pi^2 \sim R_\mrm{c}^2$
for renormalization scales such that $L_1 \sim k_\mrm{hi}$. 
However, the crucial difference between $\rho_\pi^2$ and $\bar{\rho}_\pi^2$
is that the latter is an observable, while the former absorbs a divergence 
and so is scheme- and regulator-dependent.

As an explicit example, let us consider the P-wave excited state of 
$\nuc{11}{Be}$ with neutron
separation energy $B_{\mrm{s}1}=0.182~\mrm{MeV}$ \cite{AjzenbergSelove:1990zh}. 
The breakdown scale for this
EFT was argued in Sec.~\ref{sec:Swavenhalo} to be $\khi\sim80~\mrm{MeV}$. 
These scales then give us the expansion parameter
$k_\mrm{lo}/k_\mrm{hi}\approx0.2$ for the P-wave system. 
The corresponding charge radius formula is simply organized as
\begin{equation}
r_\mrm{\nuc{11}{Be}^*}^2=\underbrace{\rho_\mrm{c}^2+\bar{\rho}_\pi^2}_{1/k_\mrm{hi}^2}
+\ldots
+\underbrace{\left[r_\mrm{pt,LO}^2+\ldots
\right]}_{1/(k_\mrm{lo} k_\mrm{hi}) \, \times \, 1/\Ac^2}
+\ldots
\label{eq:chargeradiusformulaneutronhaloP1}
\end{equation}
In this case, the LO result is given by the combination of the charge radius 
of $\nuc{10}{Be}$ and an undetermined short-range
parameter. The dots refer to corrections due to
non-included interactions and the finite neutron size. 
We show the point-charge contribution explicitly to emphasize 
that it appears at N$^2$LO in the heavy-core power counting. This means that 
the charge radius for the P-wave state in
$\nuc{11}{Be}$ cannot be predicted in HEFT using the heavy-core
power counting, unless the short-range parameter can be fixed to some other 
observable. 

\section{Conclusion}
\label{conclusion}
%
HEFT offers a systematic approach to make model-independent
predictions of low-energy observables.  In this paper we have
discussed a new power-counting scheme for systems with a heavy-core nucleus,
and we have derived the finite-size contributions to charge radii of
one-nucleon halos. HEFT in general is restricted by appearances of
short-range operators at rather low orders.  With the heavy-core power
counting, these restrictions are even enhanced for some systems and
observables. For one-neutron halos where the core is much
heavier than the neutron, the point-particle result for the charge
radius is demoted from leading to subleading order since the core
recoil due to the photon interaction is very small. In contrast, in the
case of an S-wave system, the LO charge radius is given by the
finite-size contributions of the constituents. For a P-wave
one-neutron halo the heavy-core version of HEFT is
non-predictive at LO, since the LO charge radius includes an
undetermined short-range operator~\footnote{As we were finalizing this manuscript we found that a similar conclusion
has been reached by Elkamhawy and Hammer~\cite{Elkamhawy:2019}.}. 

Note, however, that not all systems
are made less predictive in the heavy-core power counting. For proton
halos there are no issues for the charge-radius results due to the
core being heavy (as shown in the Appendix). This is due to the fact that the photon also couples
to the proton field, which has a larger recoil than the core
field. Furthermore, the expectation for the future is that more cluster
data will become available and that this data can then be used to fix the
parameters of the corresponding HEFT.

While we considered in detail the case of one-nucleon halo charge radii,
the suppression of some contributions by factors 
of the inverse of the number of core nucleons is not restricted to
this class of observables. The suppression for radii can
be traced to the small recoil of the core or, equivalently,
to the fact that the heavy-core propagator is static at leading order.
Similar effects will in principle be present in any calculation
at the loop level, where the propagator appears, for example
the structure (energies, form factors, {\it etc.}) of two-nucleon halos
or two-core systems.
We leave the investigation of these additional implications of heavy cores 
to the future.

\section*{Acknowledgments}
%
DRP and UvK acknowledge the hospitality of Chalmers University of Technology 
where this research was initiated. DRP thanks W. Elkamhawy for useful discussion.
ER and CF were supported by the European Research Council
under the European Community's Seventh Framework Programme
(FP7/2007-2013)/ERC grant agreement no.~240603,
by the Swedish Foundation for International Cooperation in Research and
Higher Education (STINT, IG2012-5158), and 
by the Swedish Research Council (dnr. 2010-4078).
The work of DRP was supported by the US Department of Energy under
contract DE-FG02-93ER-40756 and by the
ExtreMe Matter Institute EMMI
at the GSI Helmholtzzentrum f\"ur Schwerionenphysik, Darmstadt, Germany.
UvK's research was supported in part  
by the U.S. Department of Energy, Office of Science, 
Office of Nuclear Physics, under Award Number 
DE-FG02-04ER41338,
and
by the European Union Research and Innovation program Horizon 2020
under grant agreement no. 654002.

\appendix
\numberwithin{equation}{section}
\section{Single-proton halos
\label{sec:app_oneproton}}
%
For proton halos one needs additionally to account
for Coulomb effects. Denoting by $Z_\mrm{c}$ the charge of the core
and by $\alpha=e^2/(4\pi)$ the fine-structure constant,
the strength of the Coulomb interaction is characterized
by the momentum scale $\kC\equiv Z_\mrm{c}\alpha\mR$. 
At energy $E$ the relative importance 
of Coulomb is given by the 
Sommerfeld parameter
$\eta\equiv k_\mrm{C}/\sqrt{2\mR E}$.
For moderate $Z_\mrm{c}$, as in light nuclei,
we expect $k_\mrm{C}\simle k_\mrm{lo}$.
As $Z_\mrm{c}$ increases the Coulomb force experienced by the 
halo proton increases. In that case, it might be appropriate
to consider the limit $k_\mrm{lo}/k_\mrm{C}\ll 1$.
The effects of Coulomb in proton halo 
systems have been examined in Refs. 
\cite{Zhang:2014zsa,Lensky:2011he,Ryberg:2013iga,Ryberg:2014exa,Zhang:2015ajn,Ryberg:2015lea,Schmickler:2019ewl}.

The charge form factor for proton halos involves the coupling of the
photon to both the core and the proton.
The point-like contribution to the charge radius is therefore not kinematically
suppressed by $1/A_\mrm{c}^2$ and the scalings are naively
given by
\begin{equation}
r_\mrm{pt}^2\sim
\left\{\begin{array}{cc}1/k_\mrm{lo}^2~,\mrm{~S\text{-}wave~proton~halo,}
\\
1/(k_\mrm{lo}k_\mrm{hi})~,\mrm{~P\text{-}wave~proton~halo.}\end{array}\right.
\label{eq:naiveprotonhaloscaling}
\end{equation}
These naive scalings are only valid if the Coulomb momentum is a
low-momentum scale, that is $\kC\simle\klo$. If instead we are in the
strong Coulomb regime, $\kC\gg\klo$,
the predictive power of LO calculations is reduced, which has been analyzed 
for S-wave one-proton halo states in {R}ef.~\cite{Ryberg:2015lea, Schmickler:2019ewl}. 
For example, the point-like contribution
for S waves scales as $r_\mrm{pt}\sim1/\kC$ if $\kC\gg\klo$, which
implies that the point-like contribution becomes suppressed by the
strong Coulomb repulsion. 

\subsection{S-wave proton halos}
%
The procedure for deriving the charge-radius formula for S-wave proton
halos is similar to the neutron case; see
Refs.~\cite{Ryberg:2013iga,Ryberg:2015lea}. However, there are four
main differences: 
\begin{itemize}
\item 
[(i)]
The total charge of the system is $Z_\mrm{c}+1$. 

\item 
[(ii)]
The Coulomb interaction enters proton-core scattering and the wavefunction 
renormalization is given by
\begin{equation} 
\mathcal{Z}_0=
\frac{6\pi k_\mrm{C}}{g_0^2\mR^2}
\times\left\{\begin{array}{lc}
\left(\frac{6k_\mrm{C}^2}{\mR}\frac{\d h_0(\eta)}{\d E}\right)^{-1}
\bigg|_{E=-B_{\mrm{s}0}}\;,&\mrm{LO}\;,
\\
\left( \frac{6k_\mrm{C}^2}{\mR}\frac{\d h_0(\eta)}{\d E}-3k_\mrm{C}r_0 \right)^{-1}
\bigg|_{E=-B_{\mrm{s}0}}\;,
&\mrm{NLO}\;,\end{array}\right.
\label{eq:wfnormprotonS}
\end{equation}
where 
\begin{equation} 
h_0(\eta)=\psi(i\eta)+\frac{1}{2i\eta}-\log{(i\eta)},
\end{equation}
with $\psi$ being the polygamma function. 
For $\kC\gg\gamma_0$,
\begin{equation} 
\frac{6k_\mrm{C}^2}{\mR}\frac{\d h_0(\eta)}{\d E}
=1+\mathcal{O}\!\left(\frac{\gamma_0^2}{\kC^2}\right). 
\label{eq:wfnormClim}
\end{equation}
As before, some higher-order terms are kept at NLO in 
Eq. \eqref{eq:wfnormprotonS}.

\item 
[(iii)]
The photon couples also to the proton in the
proton-core loop of Fig.~\ref{fig:haloformfactordiagram}(c),
according to Eq. \eqref{nucleonFF}.

\item 
[(iv)]
Coulomb interactions enter the loops in Fig.~\ref{fig:haloformfactordiagram}.
The bound-state wavefunction is the Whittaker $W$-function
$W_{-\kC/\gamma_0,1/2}\left(2\gamma_0r\right)$ instead of the exponential
$\exp{(-\gamma_0r)}$.  

\end{itemize}

Taking these differences into account, the
resulting charge radius formula for an S-wave proton halo system is
\begin{equation}
r_\mrm{ch}^2=
\left.\left(\frac{6k_\mrm{C}^2}{\mR}
\frac{\d h_0(\eta)}{\d E}-3k_\mrm{C}r_0\right)^{-1}\right|_{E=-B_{\mrm{s}0}}
\left[r_\mrm{pt,LO}^2
+\frac{Z_\mrm{c}\rho_\mrm{c}^2+\rho_\mrm{p}^2}{Z_\mrm{c}+1}
-3\kC r_0\rho_\sigma^2\right]
+\ldots~.
\label{eq:chargeradiusformulaprotonSwave}
\end{equation}
The leading-order point-charge contribution without the effective-range correction is given by
\begin{equation}
r_\mrm{pt,LO}^2=
\frac{6\pi k_\mrm{C}}{g_0^2\mR^2}
\frac{\Gamma''_\mrm{loop}(0)}{2e(Z_\mrm{c}+1)}~,
\label{eq:pointprotonS}
\end{equation}
where the loop-diagram is given in {R}ef.~\cite{Ryberg:2015lea} as
\begin{align}
\Gamma_\mrm{loop}(\QQ)=&-eZ_\mrm{c} 
\frac{g_0^2\mR^2}{8\pi^4}
\, \Gamma\!\left(1+\kC/\gamma_0\right)^2
\int\!\d r \, j_0\!\left(f Q r\right) \, 
W_{-\kC/\gamma_0,1/2}\!\left(2\gamma_0r\right)^2
\nonumber\\
&+\left[\left(f\to1-f\right),~\left(Z_\mrm{c}\to1\right)\right]~,
\label{eq:loopdiagramprotonS}
\end{align}
where $j_0$ is a spherical Bessel function of the first kind.

The $1/2^+$ excited state of $\nuc{17}{F}$ was considered in
HEFT by Ryberg {\it et al.}~\cite{Ryberg:2013iga,Ryberg:2015lea}. 
The $1/2^+$ excited state is located at $0.105~\mrm{MeV}$ below
threshold~\cite{Tilley:1993zz},
which then defines the low-momentum scale
$\klo\sim14~\mrm{MeV}$. The first excitation of the $\nuc{16}{O}$ core
is at about $6~\mrm{MeV}$~\cite{Tilley:1993zz}, 
so the size of the core defines the
breakdown scale $\khi$ of about $1/R_\mrm{c}\sim60$--$70~\mrm{MeV}$, giving
an expansion parameter $\klo/\khi\sim0.2$. However, for the
$\nuc{16}{O}$--proton system the Coulomb momentum scale
$\kC=51.2~\mrm{MeV}$ is much larger than the low-momentum scale and
$3\kC r_0$ is very close to unity~\cite{Ryberg:2015lea}. 
This makes the effective-range
prefactor in Eq.~\eqref{eq:chargeradiusformulaprotonSwave} very large,
so this proton halo state cannot be well
described without the inclusion of effective-range
corrections.
In practice, the $\rho_\sigma^2$ counterterm then enters at the same order as the finite-size
contributions. Furthermore, in this strong Coulomb regime
the LO point-like charge radius contribution, $r_\mrm{pt,LO}$, scales with 
$1/\kC$, as was discussed in {R}ef.~\cite{Ryberg:2015lea}. 
Therefore, organizing the charge-radius formula for the S-wave proton halo at 
hand, we have
\begin{equation}
r_\mrm{\nuc{17}{F}^*}^2=\frac{1}{1-3\kC r_0}
\bigg[\underbrace{r_\mrm{pt,LO}^2}_{1/\kC^2}
+\underbrace{\frac{Z_\mrm{c}}{Z_\mrm{c}+1}\rho_\mrm{c}^2
  -3k_\mrm{C}r_0\rho_\sigma^2}_{1/\khi^2}\bigg]+\ldots~,
\label{eq:rch17f}
\end{equation}
since $3\kC r_0\sim1$ and $\kC\gg\gamma_0$ for $\nuc{17}{F}^*$. 

The point-like contribution to the charge radius of $\nuc{17}{F}^*$, including the
finite-range correction, was evaluated by 
Ryberg \textit{et al.}~\cite{Ryberg:2015lea} to $r_\mrm{pt,LO}^2 / \left( 1-3\kC
  r_0\right) = \left( 2.20 \pm 0.11 \mathrm{~fm} \right)^2$. Here, the assigned uncertainty originates
from the asymptotic normalization coefficient (ANC) obtained
from a fit to proton radiative-capture on \nuc{16}{O}. One should note
that corrections from finite-size contributions and the short-range
operator can be as large as $\kC/\khi \approx 70\%$, unless at least parts of
those corrections can be resummed.
But Eq.~\eqref{eq:rch17f} means that---at least formally---one may not
add the finite-size contributions to the charge-radius result of
$\nuc{17}{F}^*$ without also including the unknown short-range
parameter $\rho_\sigma^2$.
%
\subsection{P-wave proton halos}
%
As for the S-wave proton halos, many of the details in the derivation
of charge-radius formula for a P-wave proton halo are the same as for
the P-wave neutron halo. Again, the differences are that the photon
couples also to the proton, the bound-state wavefunctions are
Whittaker W-functions, the total charge is $Z_\mrm{c}+1$ and the wavefunction 
renormalization is now given by
\begin{equation}
\mathcal{Z}_1=\frac{6\pi}{g_1^2\mR^2}
\left(r_1-\frac{2k_\mrm{C}}{\mR}\frac{\d}{\d E}h_1(\eta)\right)^{-1}
\bigg|_{E=-B_{\mrm{s}1}}~,
\end{equation}
with $h_1(\eta)=p^2(1+\eta^2)h_0(\eta)$. For details, see
Refs.~\cite{Zhang:2014zsa,Ryberg:2014exa}. 
The resulting charge-radius formula for a P-wave proton halo system is
\begin{equation}
r_\mrm{ch}^2=\underbrace{r_\mrm{pt,LO}^2}_{1/(\klo\khi)}
+\underbrace{\frac{Z_\mrm{c}}{Z_\mrm{c} +1} \rho_\mrm{c}^2 + \bar{\rho}_\pi^2}_{1/\khi^2}+\dots~,
\label{eq:chargeradiusformulaprotonPwave}
\end{equation}
where $\bar{\rho}_\pi^2$ 
is the renormalized short-range parameter.
Here we emphasize the contribution $r_\mrm{pt,LO}^2$ is not kinematically
suppressed for systems where both constituents are charged, since the photon couples
through minimal substitution also to the proton. Therefore,
if $\bar{\rho}_\pi^2$ is subleading to $r_\mrm{pt,LO}^2$, it
is possible to give a low-order charge-radius prediction of P-wave
proton halos without including the short-range parameter
$\bar{\rho}_\pi^2$. 

The point-charge contribution $r_\mrm{pt,LO}^2$ was derived in 
{R}ef.~\cite{Ryberg:2014exa} and it is given by
\begin{equation}
r_\mrm{pt,LO}^2=-\frac{3\mathcal{Z}_1}{e(Z_\mrm{c}+1)}\Gamma''_\mrm{loop}(0)~,
\end{equation}
with the loop-diagram given by
\begin{align}
\Gamma_\mrm{loop}(Q)=&-\frac{e(Z_\mrm{c}+1)g_1^2\mR^2\Gamma
\left(2+\kC/\gamma_1\right)^2\gamma_1^2}{3\pi}
\nonumber\\
&\times\int\d r \left[1-\left((1-f)^2+Z_\mrm{c}f^2\right)
\frac{r^2Q^2}{6(Z_\mrm{c}+1)}+\mathcal{O}\left(Q^4\right)\right]
W_{-\kC/\gamma_1,3/2}(2\gamma_1r)^2~.
\end{align}

The one-proton separation energy of
$\nuc{8}{B}$ is $B_{\mrm{s}1}\simeq 0.138~\mrm{MeV}$~\cite{TILLEY2004155},  giving a low-momentum scale
$\klo\sim15~\mrm{MeV}$. The $\nuc{7}{Be}$ core has two low-lying
states that were both included explicitly into the field theory: the
$3/2^-$ ground state and the $1/2^-$ excited state at
$0.429~\mrm{MeV}$~\cite{Tilley:2002vg}. It can be argued that the breakdown scale for the
EFT is given by the alpha-particle threshold at a momentum scale of
$k_\alpha\simeq 51~\mrm{MeV}$~\cite{Tilley:2002vg} and thus the expansion parameter is about
$\klo/\khi\sim0.3$, or even as large as $\kC/\khi\sim0.5$ with $\kC =
23.8$ MeV.

The leading-order contribution to the charge radius of \nuc{8}{B} was
evaluated by Ryberg {\it et al.}~\cite{Ryberg:2014exa} to
$r_\mrm{pt,LO}^2 = \left( 2.56 \pm 0.08 \mathrm{~fm}
\right)^2$. Again, the uncertainty estimate comes from the relevant
ANCs, which in this case were adopted from a microscopic \textit{ab initio}
computation by Nollett and Wiringa~\cite{Nollett:2011}. Alternatively,
one can obtain the ANCs from a fit to proton radiative-capture
$\nuc{7}{Be}(p,\gamma)\nuc{8}{B}$. Such a fit was performed by Zhang
{\it et al.}~\cite{Zhang:2014zsa} and the ANCs are very
consistent with the computed ones.
However, the large expansion parameter suggests the finite-size and
(unknown) short-range contributions to the charge radius of \nuc{8}{B} can be
significant.
\bibliography{refs-heavyhalo}
\bibliographystyle{elsarticle-num}

\end{document}